\documentstyle[epsf]{ioplppt}
\begin{document}
\jl{6}

\title{
  	The complete spectrum of the area from recoupling theory 	
	in loop quantum gravity 
       }

\author{Simonetta Frittelli\footnote{e-mail: simo@artemis.phyast.pitt.edu},
	Luis Lehner\footnote{e-mail: luisl@artemis.phyast.pitt.edu} and 
	Carlo Rovelli\footnote{e-mail: rovelli@pitt.edu}}
\address{
 	Department of Physics and Astronomy, University of 	
	Pittsburgh, Pittsburgh, PA 15260, USA.
	}
\date{May 13, 1996}
\maketitle

\begin{abstract}

 We compute the complete spectrum of the area operator in the
 loop representation of quantum gravity, using recoupling theory.   
 This result extends previous derivations, which did not include the 
 ``degenerate'' sector, and agrees with the recently computed 
 spectrum of the connection-representation area operator. 

\end{abstract}


\newcommand{\bra}[1]{\Bigg{\langle} \!\!\!\!#1 \: \Bigg |  }
\newcommand{\brac}[1]{ \langle \  #1 \: |  }
\newcommand{\ket}[1]{\left| #1 \right\rangle }
\newcommand{\LEFT}{{\big\langle}}
\newcommand{\RIGHT}{{\big\rangle}}
\newcommand{\Bx}{{\bbox{x}}}

\def\SSs#1{{\scriptscriptstyle #1}}
\def\Ss#1{{\scriptstyle #1}}
\def\Ds#1{{\displaystyle #1}}


\newcommand{\SixBox}[6]{
\begin{array}{ccc} #1 & #2 & #3 \\
                   #4 & #5 & #6  
\end{array}}
\newcommand{\NineBox}[9]{
\begin{array}{ccc} #1 & #2 & #3 \\
                   #4 & #5 & #6 \\
                   #7 & #8 & #9  
\end{array}}

\def\SixJ[#1,#2,#3;#4,#5,#6]{
{\left\{\SixBox{#1}{#2}{#3}{#4}{#5}{#6}\right\}}}
\def\Tet[#1,#2,#3;#4,#5,#6]{
{Tet\left[\SixBox{#1}{#2}{#3}{#4}{#5}{#6}\right]}}
\def\NineJ#1,#2,#3;#4,#5,#6;#7,#8,#9]{
{\left\{\NineBox{#1}{#2}{#3}{#4}{#5}{#6}{#7}{#8}{#9}\right\}}}

\def\PBOX[#1]#2{\mbox{\setlength{\unitlength}{#1 pt} #2}}
\def\CBOX[#1]#2{\begin{array}{c} \PBOX[#1]{#2} \end{array}}
\def\MPIC[#1,#2]#3{{\begin{picture}(#1,#2) {#3} \end{picture}}}


\def\RECOUPLINGuno#1#2#3#4#5{{\MPIC[50,40]{
          \put( 0,0){$#2$}\put( 0,30){$#3$}
          \put(45,0){$#5$}\put(45,30){$#4$}
          \put(10,10){\line(1,1){10}}\put(10,30){\line(1,-1){10}}
          \put(30,20){\line(1,1){10}}\put(30,20){\line(1,-1){10}}
          \put(20,20){\line(1,0){10}}\put(22,25){$#1$}
          \put(20,20){\circle*{3}}\put(30,20){\circle*{3}}
             }}}
\def\RECOUPLINGdue#1#2#3#4#5{{\MPIC[40,50]{
          \put( 0,0){$#2$}\put( 0,40){$#3$}
          \put(35,0){$#5$}\put(35,40){$#4$}
          \put(10,10){\line(1,1){10}}\put(10,40){\line(1,-1){10}}
          \put(20,30){\line(1,1){10}}\put(20,20){\line(1,-1){10}}
          \put(20,20){\line(0,1){10}}\put(22,22){$#1$}
          \put(20,20){\circle*{3}}\put(20,30){\circle*{3}}
             }}}


\def\FIGSimmetrizer#1{{\MPIC[35,25]{{
        \put(15, 0){\line(0,1){10}}\put(20, 0){\line(0,1){10}}
        \put(15, 0){\line(1,0){ 5}}\put(15,10){\line(1,0){ 5}}
        \put(15,25){\line(1,0){5}}  \put(15,15){$#1$}
        \put(15,15){\oval(30,20)[l]}\put(20,15){\oval(30,20)[r]} 
}}}}


\def\FIGThreeVertex#1#2#3{{\MPIC[40,40]{{
       \put(15,15){\line(-1, 1){10}} \put( 4,27){$#1$}
       \put(15,15){\line( 1, 1){10}} \put(22,27){$#2$}
       \put(15, 5){\line(0,1){10}}   \put(17,1){$#3$}
       \put(15,15){\circle*{3}}
}}}}

\def\FIGvertexExchange{{\MPIC[40,50]{{
       \put(30,25){\line(-3, 2){30}} \put( 0,35){$a$}
       \put( 0,25){\line( 3, 2){13}} 
       \put(30,45){\line(-3,-2){13}} \put(30,35){$b$}
       \put(15,25){\oval(30,20)[b]}  \put(15,15){\circle*{3}}
       \put(15, 5){\line(0,1){10}}   \put(17,1){$c$}
}}}}


\def\FIGThetaNet#1#2#3{{\MPIC[40,40]{{
        \put(18,32){$#1$}
        \put( 0,15){\line(1,0){40}} \put(18,17){$#2$}
        \put(20,15){\oval(40,30)}   \put(18, 2){$#3$}
        \put( 0,15){\circle*{3}}    \put(40,15){\circle*{3}}
}}}}


\def\FIGThetraedron#1#2#3#4#5#6{{\MPIC[55,40]{{
        \put( 0,15){\line(1,-1){15}} \put(0,22){$\Ss{#2}$}
        \put( 0,15){\line(1, 1){15}} \put(0, 0){$\Ss{#1}$}
        \put( 0,15){\circle*{3}}
        \put(30,15){\line(-1, 1){15}} \put(28,20){$\Ss{#4}$}
        \put(30,15){\line(-1,-1){15}} \put(28, 2){$\Ss{#5}$}
        \put(30,15){\circle*{3}}
        \put( 0,15){\line(1,0){30}} \put(12,16){$\Ss{#6}$}
        \put(15,30){\line(1,0){25}} \put(15,30){\circle*{3}}
        \put(15, 0){\line(1,0){25}} \put(15, 0){\circle*{3}}
        \put(40, 0){\line(0,1){30}} \put(42,12){$\Ss{#3}$}
}}}}



\def\stat{{\MPIC[15,20]{{
  \put(9.5,14){\line(1,0){5.1}}
  \put(15,14){\line(1,0){1}}
  \put(17,14){\line(1,0){1}}
  \put(6,14){\line(1,0){1}}
  \put(8,14){\line(1,0){1}}
  \put(9.5,6){\line(1,0){5.1}}
  \put(15,6){\line(1,0){1}}
  \put(17,6){\line(1,0){1}}
  \put(6,6){\line(1,0){1}}
  \put(8,6){\line(1,0){1}}
  \put(11.5,12){${\scriptscriptstyle p} $}
  \put(11.5,7.){${\scriptscriptstyle q}$}
}}}}

\def\Tstat{{\MPIC[15,20]{{
  \put(9.5,14){\line(1,0){5.1}}
  \put(15,14){\line(1,0){1}}
  \put(17,14){\line(1,0){1}}
  \put(6,14){\line(1,0){1}}
  \put(8,14){\line(1,0){1}}
  \put(9.5,6){\line(1,0){5.1}}
  \put(15,6){\line(1,0){1}}
  \put(17,6){\line(1,0){1}}
  \put(6,6){\line(1,0){1}}
  \put(8,6){\line(1,0){1}}
  \put(12,6){\line(0,1){8}}
  \put(13.5,10){${\scriptscriptstyle 2} $}
  \put(9.1,12.2){${\scriptscriptstyle p} $}
  \put(9.1,6.8){${\scriptscriptstyle q}$}
  \put(12,6){\circle*{1}}
  \put(12,14){\circle*{1}}
  \put(11.2,14.9){$ \mbox{\boldmath{$\scriptscriptstyle x$}}$ }
  \put(11.2,4.){$ \mbox{\boldmath{$\scriptscriptstyle y$}} $}
}}}}


\def\trip{{\MPIC[15,20]{{
  \put(10,10){\line(0,-1){5}}
  \put(10,10){\line(1,1){4}}
  \put(10,10){\line(-1,1){4}}
  \put(8,12){\line(-1,0){3.5}}
  \put(7.5,13.5){$\SSs{p}$}
  \put(12,13.5){$\SSs{q}$}
  \put(10,5.5){$\SSs{r}$}
  \put(5.2,12.5){$\SSs{2}$}
  \put(8,12){\circle*{1}}
}}}}

\def\triq{{\MPIC[15,20]{{
  \put(10,10){\line(0,-1){5}}
  \put(10,10){\line(1,1){4}}
  \put(10,10){\line(-1,1){4}}
  \put(12.5,12.5){\line(-1,0){3}}
  \put(7.5,13.5){$\SSs{p}$}
  \put(12,13.5){$\SSs{q}$}
  \put(10,5.5){$\SSs{r}$}
  \put(10,12.8){$\SSs{2}$}
  \put(12.5,12.5){\circle*{1}}
}}}}

\def\trir{{\MPIC[15,20]{{
  \put(10,10){\line(0,-1){5}}
  \put(10,10){\line(1,1){4}}
  \put(10,10){\line(-1,1){4}}
  \put(10,8){\line(-1,0){3.5}}
  \put(7.5,13.5){$\SSs{p}$}
  \put(12,13.5){$\SSs{q}$}
  \put(10,5.5){$\SSs{r}$}
  \put(8,8.8){$\SSs{2}$}
  \put(10,8){\circle*{1}}
}}}}


\def\normgras{{\MPIC[15,20]{{
    \put(8,8){\line(2,-1){7}}
  \put(8,8){\line(-3,5){4.5}}
  \put(6.8,10){\line(1,0){2.2}}
  \put(1.8,10.5){$\Ss{p}$}
  \put(12,3.){$\Ss{q}$}
  \put(11,10){$\Ss{\scriptstyle{2}}$}
  \put(8,8){\circle*{1.5}}
  \put(6,6.){$\Ss{i}$}
  \put(9,8){\oval(4,4)[tr]}
  \put(11,8){\line(0,-1){1.5}}
  \put(6.8,10){\circle*{1}}
  \put(11,6.5){\circle*{1}}
}}}}

\def\Terc{{\MPIC[15,20]{{
  \put(6,2){\line(0,1){16}}
  \put(6,10){\line(1,0){8}}
  \put(4,2){$\Ss{q}$}
  \put(4,18){$\Ss{p}$}
  \put(10,10.2){${}^r$}
  \put(4.8,12){$\Ss{p}$}
  \put(4.8,8){$\Ss{q}$}
  \put(6,10){\circle*{1.5}}
  \put(6,10){\oval(8,8)[l]}
  \put(0,9.5){${}^2$}
  \put(6,14){\circle*{1}}
  \put(6,6){\circle*{1}}
}}}}

\def\Prim{{\MPIC[15,20]{{
  \put(6,0){\line(0,1){19}}
  \put(6,10){\line(1,0){8}}
  \put(4,5){$\Ss{q}$}
  \put(4,19){$\Ss{p}$}
  \put(13,10.2){${}^r$}
  \put(4.8,15){$\Ss{p}$}
  \put(4,11.5){$\Ss{p}$}
  \put(6,10){\circle*{1}}
  \put(6,15){\oval(4,4)[l]}
  \put(2.5,13){${}^2$}
  \put(6,17){\circle*{1}}
  \put(6,13){\circle*{1}}
}}}}

\def\Threelin{{\MPIC[15,20]{{
  \put(6,0){\line(0,1){16}}
  \put(6,8){\line(1,0){10}}
  \put(2,2){$\Ss{q}$}
  \put(2,14){$\Ss{p}$}
  \put(10,8){${}^r$}
  \put(6,8){\circle*{2}}
}}}}

\def\PossiB{{\MPIC[15,20]{{
  \put(6,0){\line(0,1){20}}
  \put(6,10){\line(1,0){8}}
  \put(4,1){$\Ss{q}$}
  \put(4,15){$\Ss{p}$}
  \put(10,10.2){${}^r$}
  \put(4.8,5){$\Ss{q}$}
  \put(4,8.5){$\Ss{q}$}
  \put(6,10){\circle*{1}}
  \put(6,5){\oval(4,4)[l]}
  \put(2.5,4){${}^2$}
  \put(6,7){\circle*{1}}
  \put(6,3){\circle*{1}}
}}}}

\def\PossiA{{\MPIC[15,20]{{
  \put(6,0){\line(0,1){20}}
  \put(6,10){\line(1,0){8}}
  \put(4,5){$\Ss{q}$}
  \put(4,19){$\Ss{p}$}
  \put(10,10.2){${}^r$}
  \put(4.8,15){$\Ss{p}$}
  \put(4,11.5){$\Ss{p}$}
  \put(6,10){\circle*{1}}
  \put(6,15){\oval(4,4)[l]}
  \put(2.5,13){${}^2$}
  \put(6,17){\circle*{1}}
  \put(6,13){\circle*{1}}
}}}}

\def\PossiC{{\MPIC[15,20]{{
  \put(6,0){\line(0,1){20}}
  \put(6,10){\line(1,0){8}}
  \put(4,2){$\Ss{q}$}
  \put(4,18){$\Ss{p}$}
  \put(10,10.2){${}^r$}
  \put(4.8,12){$\Ss{p}$}
  \put(4.8,8){$\Ss{q}$}
  \put(6,10){\circle*{1}}
  \put(6,10){\oval(8,8)[l]}
  \put(0,9.5){${}^2$}
  \put(6,14){\circle*{1}}
  \put(6,6){\circle*{1}}
}}}}



\section{Introduction}

One of the central results of the loop approach to quantum gravity
\cite{loops} is the derivation of discrete properties of spacetime
geometry
\cite{weave,eigenvalues,prediction,depietri-rovelli96,ashtekar-lewandowski}.
This derivation realizes the old idea \cite{garay} that spacetime might
exhibit some kind of quantum discreteness at the Planck scale
\cite{isham}.  One of the manifestations of such discreteness is the
fact that the operators associated to physical area and volume have
discrete spectra~\cite{eigenvalues}.  This fact leads to the prediction
that measurements of areas and volumes at the Planck scale would yield
quantized values \cite{prediction}.   The explicit computation of the
spectra of area and volume is thus a relevant step towards
understanding the physics of the quantum gravitational field.  Partial
results on these spectra, for instance, have already been employed in
discussing quantum gravitational corrections of black hole
radiation~\cite{cr1} and black hole entropy~\cite{cr2}.

Here, we consider the area operator.  A regularization technique for
the definition of this operator was introduced in~\cite{weave}, where
some of the eigenvalues were computed.  A more complete treatment was
given in~\cite{eigenvalues}, where the spectrum was computed in full
except for a ``degenerate'' sector formed  by the states in which
vertices or edges of the spin network lie on the surface.  It is
difficult to treat this degenerate sector --whose existence was pointed
out by A Ashtekar-- using the original regularization, because
additional divergences appear.  As a result, earlier works on the
calculation of geometry eigenvalues in the loop representation exhibit
incomplete spectra for the area operator.  In this paper we introduce
an alternative regularization, whose action is well defined on every
loop state.  This is done in Section~\ref{II}.   This regularization
allows us to compute the full spectrum of the area operator.  The
computation is performed in Section~\ref{III}, using recoupling theory,
which provides a powerful computation technique in quantum
gravity~\cite{depietri-rovelli96}.   Our calculation here is a further
proof of the effectiveness of recoupling theory in this context.

The complete spectrum of the area has been recently computed in
Ref.~\cite{ashtekar-lewandowski} in the connection representation of
quantum gravity~\cite{connection}.  The spectrum we obtain here fully
agrees with the one given in Ref.~\cite{ashtekar-lewandowski}.  Notice
that the two representations are unitarily isomorphic, as argued by
Lewandowski \cite{lew}, and, recently, by DePietri \cite{dp}, but the
regularization techniques employed to define physical operators differ,
so that the agreement is a non-trivial result.

  \section{The area operator			\label{II}}

  \noindent 
Consider a two-dimensional surface $\Sigma$ embedded in a
three-dimensional metric manifold $M$.  Let $\sigma^u$, $u=1,2$, be
coordinates on $\Sigma$, $x^a$, $a=1,2,3$, coordinates on $M$, and let
$x^a(\sigma^u)$ represent the embedding.  If we represent the metric of
$ M$ by means of the inverse densitized triad $\tilde E^{ai}$, the area
of $\Sigma$ is given by
\begin{equation}
	A[\Sigma] =   \int_\Sigma 
			d^2\sigma \, 
			     \sqrt{\tilde{E}^{ai}\tilde{E}^b_i\, 
	   		    n_a           n_b
				   }			\label{clarea}
  \end{equation}
where $n_a=\frac{1}{2} \epsilon_{abc} \frac{\partial x^b}{\partial
\sigma^1} \frac{\partial x^c}{\partial \sigma^2}$ is the one-form
normal to $\Sigma$.  Since the metric of physical space is the
gravitational field, $A[\Sigma]$ is a function of the gravitational
field: it is then represented by a quantum operator $\hat A[\Sigma]$ in
a quantum description of gravity.    In the loop quantization of
general relativity, one constructs $\hat A[\Sigma]$ by expressing the
area in terms of loop variables and then replacing the loop variables
with the corresponding quantum loop operators.   Since (\ref{clarea})
involves products of two triads $\tilde{E}^a$, the area is expected to
be a function of the loop variable of order 2 (with two ``hands'')
  \begin{equation}
	{\cal T}^{ab}[  \alpha](\mbox{\boldmath $x$},
			        \mbox{\boldmath $y$})		=
       - \mbox{Tr}   [
		      U_\alpha (\mbox{\boldmath $y$},
			        \mbox{\boldmath $x$})
	\tilde{E}^a            (\mbox{\boldmath $x$})
		      U_\alpha (\mbox{\boldmath $x$},
			        \mbox{\boldmath $y$})
	\tilde{E}^b            (\mbox{\boldmath $y$})
			      				]
  \end{equation}
where $\alpha$ denotes a loop and $U_\alpha(\mbox{\boldmath
$x$},\mbox{\boldmath $y$})$ is the parallel propagator of the Ashtekar
connection along $\alpha$ with end points at {\boldmath $x$} and
{\boldmath $y$}.    Throughout this work, we assume familiarity with
the loop representation~\cite{loops}; in particular, we refer reader
to~\cite{depietri-rovelli96} for notations and conventions. (Notice
that we deal here with conventional loop quantum gravity, not with its
quantum SU(2) deformations~\cite{lee}.)

A direct change of variables $\tilde{E}^a \,\tilde{E}^b \rightarrow
{\cal T}^{ab}$ leads to an expression which is not suitable for
quantization, due to the the presence of products of operators at the
same point. Therefore a regularization procedure is needed in order to
define $\hat  A[\Sigma]$.  This can be done by selecting a sequence of
quantities $A_\epsilon[\Sigma]$ converging to (\ref{clarea}) when
$\epsilon$ goes to zero, each of which does not involve products of
variables at the same point.  The sequence can be quantized by
substituting the dynamical  variables with the corresponding quantum
operators. The operator $\hat{A}(\Sigma)$ is then defined as a suitable
operator limit of the resulting sequence of operators.  There is a
degree of arbitrariness in any regularization procedure (in
conventional field theory: dimensional regularization, point splitting,
Pauli-Villar...).  In the present context, the regularization procedure
must satisfy two requirements. First, the classical regularized
expression must converge to the correct classical quantity (the area)
when the regulator is removed; second, the quantum operator must be
well defined in the limit and must respect the invariances of the
theory; in particular, the operator should transform correctly under
diffeomorphisms.  Contrary to a possible impression that the choice of
a regularization leaves great arbitrariness, implementing both
requirements is actually far from trivial.  The modification of the
regularization procedure for the area operator considered in this paper
is forced by the realization that the regularization considered in
\cite{eigenvalues} fails to satisfy the second viability criterion, if
one relaxes the simplifying choice of neglecting certain states.  In
the following, we construct one such regularization of the area
operator.

Let us begin by introducing a smooth coordinate $\tau$ over a finite
neighborhood of $\Sigma$, in such a way that $\Sigma$ is given by
$\tau=0$.  Consider then the three-dimensional region around $\Sigma$
defined by $-\delta/2 \le \tau \le \delta/2 $.   Partition this region
into a number of blocks $\cal D_I$ of coordinate height $\delta$ and
square horizontal section of coordinate side $\epsilon$.   For each
fixed choice of $\epsilon$ and $\delta$, we label the blocks by an
index $I$.    Later, we will send both $\delta $ and $\epsilon$ to
zero. In order to have a one-parameter sequence, we now choose $\delta$
as a fixed function of $\epsilon$.  For technical reasons, the height
of the block $\cal D_I$ must decrease more rapidly than $\epsilon$ in
the limit; thus, we put $\delta=\epsilon^k$ with any $k$ greater than
1 and smaller than 2.

Consider one of the blocks.  The intersection of the block and a $\tau
= constant$ surface is a square surface: let $A_{I}(\tau)$ be the area
of such surface.   Let $A_{I\epsilon}$ be the average over $\tau$ of
the areas of the surfaces in the block. Namely
  \begin{equation}
	A_{I\epsilon} \equiv \frac1\delta
				   \int^{\delta/2}_{-\delta/2} 
	A_{I}(\tau) 					d\tau
	= 
	\frac1\delta
	\int_{{\cal D_I}} d^3\!x\;
	      \sqrt{\tilde{E}^{ai}\tilde{E}^b_i
			    n_a           n_b	}	\;\;.
				\label{iterms}
  \end{equation} 
Summing over the blocks yields the average of the areas of the
$\tau=constant$ surfaces, and as $\epsilon$ (and therefore $\delta$)
approaches zero, the sum converges to the area of the surface $\Sigma$.
Therefore we have
  \begin{equation}
	A[\Sigma] =      \lim_{\epsilon \to 0} \sum_I 
						 A_{I\epsilon}
		  \equiv \lim_{\epsilon \to 0}   
						 A_{\epsilon}[\Sigma]. 
        				             \label{limarea}
  \end{equation}
The quantity $A_{I\epsilon}$ associated to each block can be expressed
in terms of the fundamental loop variables as follows. First, let us
pick an arbitrary fiducial background coordinate system.  For every two
points  $\mbox{\boldmath $x$}$ and $\mbox{\boldmath $y$}$, let $\alpha$
be a loop determined by $\mbox{\boldmath $x$}$ and $\mbox{\boldmath
$y$}$ -- say as the zero-area loop obtained by following back and forth
a straight segment between $\mbox{\boldmath $x$}$ and $\mbox{\boldmath
$y$}$, in the fiducial coordinate system. Then, we write
  \begin{equation}
 		A_{I\epsilon} 		  =   
	 \sqrt{ A_{I\epsilon}^2 }	\label{termarea}
  \end{equation}
and notice that
  \begin{equation}
	        A_{I\epsilon}^2 		  = 
					\frac{1}{2\delta^2}
		                        \int_{{\cal D}\otimes{\cal D}}
					d^3x\, d^3y\;
		  n_a(\mbox{\boldmath $x$})
		  n_b(\mbox{\boldmath $y$})
		    {\cal T}^{ab}
	[\alpha]
     			 (\mbox{\boldmath $x$},
			  \mbox{\boldmath $y$})			\; + \;
	O(\epsilon^5)		     .		\label{squarea}
  \end{equation}
Equation~(\ref{squarea}) holds because of the following. We have
  \begin{equation}
	{\cal T}^{ab}[\alpha]
			 (\mbox{\boldmath $x$},
			  \mbox{\boldmath $y$})	 
      = 
	2 \tilde{E}^{ai} (\mbox{\boldmath $x$}_I)
	  \tilde{E}^{b}_i(\mbox{\boldmath $x$}_I) + O(\epsilon)
							\label{2Tlimit}
 \end{equation}
for any three points {\boldmath $x$}, {\boldmath $y$}, and
$\mbox{\boldmath $x$}_I$, in ${\cal D}$. It follows that
  \begin{eqnarray}
	\fl
         \epsilon^4\,
	                     n_a    (\mbox{\boldmath $x$}_I)
			     n_b    (\mbox{\boldmath $x$}_I)
		     \tilde{E}^{ai} (\mbox{\boldmath $x$}_I)
	             \tilde{E}^b_i  (\mbox{\boldmath $x$}_I)	\nonumber\\	
	\lo=  
     \frac{1}{2\delta^2}
	\int_{{\cal D}\otimes{\cal D}} d^3\!x\,d^3\!y\;
		n_a           (\mbox{\boldmath $x$})
	        n_b           (\mbox{\boldmath $y$})
	 {\cal T}^{ab}[\alpha]
			      (\mbox{\boldmath $x$},
			       \mbox{\boldmath $y$}) 	
							\; + \;
	  O(\epsilon^5)  .					\label{ee}
\end{eqnarray}
Since
  \begin{eqnarray}
  	A_{I\epsilon      }^2 &=& \bigg(\frac1\delta
				   \int^{\delta/2}_{-\delta/2} 
	A_{I\epsilon}(\tau) 					d\tau
				\bigg)^2 \nonumber \\
			      &=& \bigg(
	\frac1\delta
	\int_{{\cal D}} d^3\!x\;
	      \sqrt{\tilde{E}^{ai}\tilde{E}^b_i
			    n_a           n_b	}
				\bigg)^2 \nonumber \\
			      &=&
         \epsilon^4\,
	                     n_a    (\mbox{\boldmath $x$}_I)
			     n_b    (\mbox{\boldmath $x$}_I)
		     \tilde{E}^{ai} (\mbox{\boldmath $x$}_I)
	             \tilde{E}^b_i  (\mbox{\boldmath $x$}_I)	\;+\;
	O(\epsilon^5),
  \end{eqnarray}
Equation~(\ref{squarea}) follows. 

Equations (\ref{limarea}), (\ref{termarea}) and (\ref{squarea}) define
our regularization for the area.  The operator $\hat{A}[\Sigma]$ can
now be defined as
  \begin{eqnarray}
		\hat A[\Sigma] 					  
								  \equiv	
 		\lim_{\epsilon\to 0} 
		\hat A_\epsilon[\Sigma],   		\label{oparea}\\
 \mbox{with}\hspace{.4cm}		
		\hat A_\epsilon[\Sigma] \label{eq:sqrt}		
								  \equiv
		\sum_I 
		\sqrt{\hat{A^2_{I\epsilon}}}, 		\label{oparea2}\\ 
\mbox{and}\hspace{.7cm}
		\hat{A^2_{I\epsilon}} 				  
								  \equiv
   \frac{1}{2\delta^2}
	\int_{{\cal D}\otimes{\cal D}} d^3\!x\,d^3\!y\;
		n_a           (\mbox{\boldmath $x$})
	        n_b           (\mbox{\boldmath $y$})
     \hat{\cal T}^{ab}[\alpha]
			      (\mbox{\boldmath $x$},
			       \mbox{\boldmath $y$}). 
	\label{oparea3}
  \end{eqnarray}
The meaning of the operator limit in (\ref{oparea}) is discussed
in~\cite{depietri-rovelli96}.

The space of states of the quantum theory is the space of loops up to
the Mandelstam relations (which essentially identify loops with the same
holonomy).  A state is usually denoted $\langle\beta|$ (for the
Mandelstam class of the loop $\beta$) or by means of a bracketed
pictorial representation of the loop $\beta$.  Any state has an
associated graph, which is characterized as a collection of edges
(smooth lines of generic multiplicity) and vertices (points at which
lines converge).  A vertex has a valence, defined as the number of
edges converging to it.  A graph is said to be $n$-valent if all its
vertices have valence $n$ or less.   A convenient basis of the space of
quantum states is the spin network basis, introduced in~\cite{spinnet}
and further discussed in~\cite{depietri-rovelli96}.  Spin network
states are given by the linear combinations of loop states obtained by
antisymmetrizing the loops along each edge of their graphs. Because of
the Mandelstam relations, these linear combinations form a complete
basis.   A spin network is characterized by an $n$-valent graph, an
assignment of an ordering to the edges converging to each vertex, a
color $p$ assigned to each edge (the number of antisymmetrized loops),
and a color $v$ assigned to each vertex (characterizing the rooting of
the loops through the vertex).  An $n$-valent spin network can be
expanded into a ``virtual'' trivalent spin network, lying in the ribbon
associated to the state --  a possibility that we exploit below.
See~\cite{depietri-rovelli96} for details.

The action of $\hat A[\Sigma]$ on the quantum states is found from the
action of the $\hat{\cal T}^{ab}[\alpha]$ operators, which we recall
here.  The operator $\hat{\cal T}^{ab}[\alpha](\mbox{\boldmath
$x$},\mbox{\boldmath $y$})$ annihilates the state $\langle\beta|$
unless the loop $\beta$ intersects the loop $\alpha$ at the points
{\boldmath $x$} {\em and\/} {\boldmath $y$}. If the loops do intersect
as needed, the action of the operator on the state gives the
(Mandelstam class of) the union of the loops $\beta$ and $\alpha$, with
two additional vertices at the points {\boldmath $x$} and {\boldmath
$y$}.  More precisely, if {\boldmath $x$} and {\boldmath $y$} fall over
two edges of $\beta$ with color $p$ and $q$ respectively, we have
  \begin{equation}
	\bra{\!\!\!\!{\CBOX[2]{\stat}}\,\,} 
 	\hat{\cal T}^{ab}[\alpha](\mbox{\boldmath $x$}, 
				  \mbox{\boldmath $y$}) 	
							=
	 - l_0^4\, p\, q\, 
		 \Delta^a[\beta,  \mbox{\boldmath $x$}] \, 
		 \Delta^b[\beta,  \mbox{\boldmath $y$}] 
	\bra{\!\!\!\!{\CBOX[2]{\Tstat}}\,\,}
\label{t2}
  \end{equation}
where 
  \begin{equation}
		\Delta^a [\beta,  \mbox{\boldmath $x$}] = 
		     \int_\beta   d\tau 
		    ~\dot{\beta}^a(\tau) 
                \delta^3 [\beta(   \tau),
				  \mbox{\boldmath $x$}].
\label{delta} 
 \end{equation}
In the picture, the loop $\alpha$ is represented as a double line
running back and forth between the intersection points {\boldmath $x$}
and {\boldmath $y$} (the two ``grasps''), hence the label 2.

\section{Spectrum of the area operator		\label{III}}
 
In this section we discuss the action of the operator $\hat A[\Sigma]$
on a generic spin network state $\brac{S}$.  Due to the limiting
procedure involved in its definition, the operator $\hat A[\Sigma]$
does not affect the graph of any state.  Furthermore, since the action
of $\hat{\cal T}^{ab}$ inside a specific coordinate block ${\cal D}$
vanishes unless the graph of the state intersects ${\cal D}$, the
action of $\hat A[\Sigma]$ ultimately consists of a sum of numerable
terms, one for each intersection $i$ of the graph with the surface.
Here we allow for spin networks having vertices on $\Sigma$ or edges
tangential to $\Sigma$, unlike previous treatments of this
problem~\cite{depietri-rovelli96}.  

Consider an intersection $i$ between the spin network and the surface.
For the purpose of this discussion, we can consider a generic point on
an edge as a ``two-valent vertex'', and thus say, without loss of
generality, that $i$ is a vertex.  In general, there will be $n$ edges
emerging from $i$.  Some of these will emerge above, some below and
some tangentially to the surface $\Sigma$.  Since we are taking the
limit in which the blocks shrink to zero, we may assume, without loss
of generality that the surface and the edges are linear around $i$ (see
below for subtleties concerning higher derivatives).  Due to the two
integrals in (\ref{oparea3}), the position of two hands of the area
operator are integrated over each block. As the action of $\hat{\cal
T}^{ab}$ is non-vanishing only when both hands fall on the spin
network, we obtain $n^2$ terms, one for every couple of grasped edges.
Consider one of these terms, in which the grasped edges have color $p$
and $q$.  The state
  \begin{equation}
		\bra{\CBOX[2]{\normgras}}_\epsilon \;  	\label{gen}
  \end{equation}
represents, up to factors, the result of the action of $\hat{\cal
T}^{ab}$ on the edges $p$ and $q$ of an $n$-valent intersection $i$.
The irrelevant edges are not shown. The edges labeled $p$ and $q$ are
generic, in the sense that their angles with the surface do not need to
be specified at this point (the two edges may also be identical).
From the definition (\ref{oparea}-\ref{oparea3}) of the area operator
and the definition (\ref{t2}-\ref{delta}) of the $\hat{\cal T}^{ab}$
operator, each term in which the grasps run over two edges of color $p$
and $q$ is of the form
 \begin{eqnarray}
	\fl
                  \frac{1}{2\delta^2}
                  \int_{{\cal D}\otimes{\cal D}} d^3\!x\, d^3\!y\; 
        n_a(\mbox{\boldmath $x$}) \Delta^a[\beta,\mbox{\boldmath $x$}] 
        n_b(\mbox{\boldmath $y$}) \Delta^b[\beta,\mbox{\boldmath $y$}]
                  p \, q \;%
                          \bra{\CBOX[2]{\normgras}}_\epsilon \nonumber\\
   \lo= \frac{1}{2\delta^2}
\int_{{\cal D}\otimes{\cal D}} \Bigg(
    n_a(\mbox{\boldmath $x$})
                 \int_\beta ds ~\dot{\beta}^a(s)
                        \delta^3 [\beta(s),\mbox{\boldmath $x$}]  \nonumber \\
    \times   n_b(\mbox{\boldmath $y$})
              \int_\beta dt ~\dot{\beta}^b(t)
                             \delta^3 [\beta(t),\mbox{\boldmath $y$}]\; 
                   p \, q \;%
                           \bra{\CBOX[2]{\normgras}}_\epsilon
			\Bigg)d^3\!x\, d^3\!y\;
                                                                \nonumber \\
  \lo=      \frac{1}{2\delta^2}
     \int_\beta ds \;  n_a(s) ~\dot{\beta}^a(s)
     \int_\beta dt \; n_b(t) ~\dot{\beta}^b(t)
                   p \, q \;
                            \bra{\CBOX[2]{\normgras}}_\epsilon
                                                               \nonumber \\
 \lo=     \frac{p \, q }{2\delta^2}
      \Bigg( \int_\beta ds \; n_a(s) ~\dot{\beta}^a(s)
        \int_\beta dt \; n_b(t) ~\dot{\beta}^b(t) 
        \Bigg)
            \bra{\CBOX[2]{\normgras}} + O(\epsilon) \; .  
                                                \label{eq:integ}
  \end{eqnarray} 
The last step in the preceding calculation is pulling the state outside
the integral sign. This is possible because the $\epsilon$-dependent
states $\bra{\CBOX[2]{\normgras}}_\epsilon$ all have the same limit
state as $\epsilon \to 0$, namely $\bra{\CBOX[2]{\normgras}}$; i.e.,
  \begin{equation}
        \bra{\CBOX[2]{\normgras}}_\epsilon = 
        \bra{\CBOX[2]{\normgras}} + O(\epsilon) 
  \end{equation} 
and, hence, the substitution of the $\epsilon$-dependent states with
their $\epsilon$-independent limit in the integral is possible up to
terms of order $O(\epsilon)$.  Note the following: 
  \begin{eqnarray} 
	\int_\beta dt \; n_b(t) ~\dot{\beta}^b(t)
     =
	\left\{ \begin{array}{l}
        0 \hspace{1cm} \mbox{if $\beta$ is tangent to $\Sigma$} \\
                                \delta/2 \hspace{1cm} \mbox{otherwise.}
                                        \end{array}
                                        \right.  
  \end{eqnarray} 
This result is independent of the angle the edge makes with the surface
because $\delta$ can always be chosen sufficiently small so that
$\beta$ crosses the top and bottom of the coordinate block $\cal D$
(this is the reason for requiring that $\delta$ goes to zero faster
than $\epsilon$).  Also, since we have chosen $k$ smaller than 2, it
follows that any edge tangential to the surface exits the box from the
side, irrespectively from its second (and higher) derivatives, for
sufficiently small $\epsilon$, and gives a vanishing contribution as
$\epsilon$ goes to zero.  Thus, the parenthetic factor in
(\ref{eq:integ}) is either $0$ or $\delta^2/4$ depending on whether one
or none of the edges lies on the surface. Consequently, in the limit
considered, the edges tangent to the surface do not contribute to the
action of the area whereas every non vanishing term takes the form
  \begin{equation}
	- \frac{ l_0^4 \,p \, q}{8} \bra{\CBOX[2]{\normgras}} . 
  \end{equation}

\begin{figure}
\epsfxsize=250pt\epsfbox{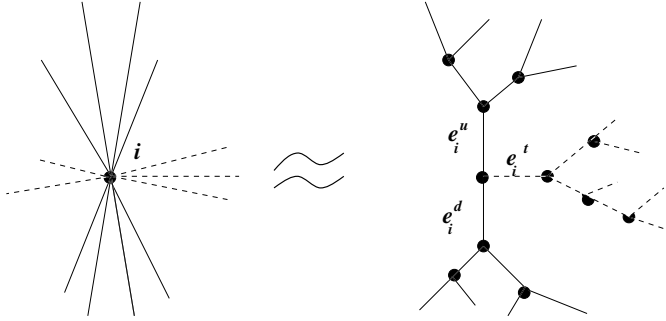}
\caption{ Trivalent expansion of an n-valent vertex. The dashed lines
indicate those lines tangent to the surface.}
\label{fig:equi}
\end{figure}

Generically, there will be several edges above, below, and tangential
to the surface $\Sigma$.  Following \cite{depietri-rovelli96}, we now
expand the vertex $i$ into a virtual trivalent spin network.  We choose
to perform the expansion in such a way that all edges above the surface
converge to a single ``principal'' virtual edge $e^u$; all edges below
the surface converge to a single principal virtual edge  $e^d$; and all
edges tangential to the surface converge to a single  principal virtual
edge $e^t$.  The three principal edges join in the principal trivalent
vertex. This trivalent expansion is shown in Figure~\ref{fig:equi}.
This choice simplifies the calculation of the action of the area, since
the sum of the grasps of one hand on {\it all\/} real edges above the
surface is equivalent to a single grasp on $e^u$ (and similarly for the
edges below the surface and $e^d$).  This follows from the identity 
  \begin{equation}
	p\!\!\!\!\!\CBOX[3]{\trip} + q\!\!\!\!\!\CBOX[3]{\triq} 
     = 	
	r\!\!\!\!\!\CBOX[3]{\trir},  			\label{eq:iden} 
  \end{equation} 
which can be proven as follows.  Using the recoupling theorem
(\ref{eq:recTheorem}), the left hand side of (\ref{eq:iden}) can be
written as 
  \begin{equation}
 	\sum_j  \Bigg( p\, \SixJ[2,p,j;q,r,p] 
		    - q\, \lambda^{2r}_j
			  \SixJ[r,p,j;q,2,q] 
		\Bigg)
        \CBOX[1]{\RECOUPLINGdue{j}{2}{p}{q}{r}} 
  \end{equation} 
where $j$ can take the values $r-2$, $r$ and $r+2$. A straightforward 
calculation making use of (\ref{eq:tetraedro}) gives 
  \begin{equation}
 	 p\, \SixJ[2,p,j;q,r,p] - q\, \lambda^{2r}_j 
             \SixJ[r,p,j;q,2,q] = r\, \delta_{jr} . 
  \end{equation} 
Thus, (\ref{eq:iden}) follows. A repeated application of the identity
(\ref{eq:iden}) allows us to slide all graspings from the real edges
down to the two virtual edges $e^u$ and $e^d$.  Thus, each intersection
contributes as a single principal trivalent vertex, regardless of its
valence. 

We are now in position to calculate the action of the area on a generic 
intersection. From the discussion above, the only relevant terms are
as follows
  \begin{eqnarray}
	\fl
 	\bra{\CBOX[2]{\Threelin}} \hat A^2_{i} 		\nonumber\\
	\lo= 
    - \frac{l_0^4}{8} \Bigg( p^2 ~ { \bra{\CBOX[3]{\PossiA}}} ~ 
      + ~ q^2 ~ 
	\bra{{\CBOX[3]{\PossiB}}}  ~ 			\nonumber\\
	+
	~ 2 \,p \, q ~  \bra{{\CBOX[3]{\PossiC}}} ~ \Bigg)
							\label{eq:arterms}
  \end{eqnarray}
where the first term comes from grasps on the edges above the surface,
the second from grasps on two edges below the surface and the third
from the terms in which one hand grasps an edge above and the other
grasps an edge below the surface.  Equation~(\ref{eq:arterms}) is an
eigenvalue problem, as it can be seen from recoupling
theory~\cite{Recoup}, since each term in the sum is proportional to the
original state (see \ref{eq:upterm}, \ref{eq:tanterm}). Therefore, we
have
  \begin{equation}
	\bra{\CBOX[2]{\Threelin}} \hat A^2_{i} 
     = - \frac{l_0^4}{8} \,( \, p^2 ~ \lambda_u ~ + ~ q^2 ~ \lambda_d  ~ +
	~ 2\, p \, q ~  \lambda_t ~ )\, \bra{\CBOX[2]{\Threelin}}. 
							\label{eq:arterms2}
  \end{equation}
The quantities $\lambda_u$, $\lambda_d$ and $\lambda_t$ are easily
obtained from the recoupling theory. Using the formulas in the
appendix, we obtain 
  \begin{equation}
 	\lambda_u  = \frac{\theta(p,p,2)}{\Delta(p)}= -\frac{(p+2)}{2p}.  
							 \label{eq:lamup}
  \end{equation}
$\lambda_d$ is obtained by replacing $p$ with $q$ in (\ref{eq:lamup}).
$\lambda_t$ has the value
  \begin{equation}
 	\lambda_t = \frac{ \Tet[p,p,r;q,q,2]}{\theta(p,q,r)} 
      = \frac{-2p(p+2) 
	-2q(q+2)+2r(r+2))}{8pq} 			 \label{eq:lamtan}. 
  \end{equation}
Substituting in (\ref{eq:arterms}), we have 
  \begin{equation}
	\bra{\CBOX[2]{\Threelin}} \hat A^2_{i}  = \frac{l_0^4}{16} 
	(\,2\,p\,(p+2)\, +\,2\,q\,(q+2) - r\,(r+2)\,) 
	\bra{\CBOX[2]{\Threelin}} 
  \end{equation}
Since $\hat A^2_{i}$ is diagonal, the square root in (\ref{eq:sqrt})
can be easily taken:
  \begin{eqnarray}
	\fl
	 \bra{\CBOX[2]{\Threelin}} \hat A_i 
	= \bra{\CBOX[2]{\Threelin}} \sqrt{\hat A^2_i}   \nonumber\\
 	\lo=\sqrt{ \frac{l_0^4}{4}~ \Bigg( 
		2\frac{p}{2}\left(\frac{p}{2}+1\right) 
 		+ 2\frac{q}{2}\left(\frac{q}{2}+1\right) - 
		\frac{r}{2}\left(\frac{r}{2}+1\right) \Bigg) } 
		\bra{\CBOX[2]{\Threelin}}
  \end{eqnarray}
Adding over the intersections and using the spin notation $\frac{p}{2}
= j^u$, $\frac{q}{2} = j^d$ and $\frac{r}{2} = j^t$, the final result
is:
  \begin{equation}
	\fl
	\brac{S}\ \hat{A}[\Sigma] = 
	\left(\frac{l^2_0}{2} \! \sum_{i\in\{S\cap\Sigma\}}\!  
	\sqrt{ 2j^u_i(j^u_i\!+\!1) + 2j^d_i(j^d_i\!+\!1) 
	- j^t_i(j^t_i\!+\!1) }\right) \  \brac{S}
  \end{equation}
This expression provides the complete spectrum of the area.  It
contains earlier results~\cite{eigenvalues} as the subset defined by
$j_i^t=0$ and $j_i^d=j_i^u$ (for every $i$).

\ack
We are very grateful to Roberto DePietri for continuous advice and
for a careful reading of the manuscript, and to Jerzy 
Lewandowski for sharing with us his results prior to publication
and for many stimulating discussions.  We wish to thank the referees
for useful criticisms.

\appendix

\section{Basic Formulae}

We present here a summary of the basic formulae of recoupling theory
(in the classical case $A=-1$ and $d=-2$) used in this work.

\noindent {(1)} The symmetrizer
\begin{equation}
\Delta_n = \CBOX[1]{\FIGSimmetrizer{n}} = (-1)^n (n+1) 
         =  (-1)^n (n+1). 
\label{eq:valSIM}
\end{equation}
\vspace{0.7cm}
\noindent {(2)} The line exchange  in a 3-Vertex
      \begin{equation}
      \CBOX[1]{\FIGvertexExchange} 
      = \lambda^{ab}_c \CBOX[1]{\FIGThreeVertex{a}{b}{c}} 
      \end{equation}
Where $\lambda^{ab}_c = (-1)^{(a+b-c)/2} ~A^{(a'+b'-c')/2}$
      and $x'=x(x+2)$.

\vspace{0.7cm}
\noindent {(3)} The evaluation of $\theta$ 
\begin{eqnarray}
\theta(a,b,c) &=&\CBOX[1]{\FIGThetaNet{a}{b}{c}} 
\label{eq:valTheta}\\
~~~ &=& \frac{ (-1)^{m+n+p} (m+n+p+1)! ~m!~n!~p!}{
                         a! ~b! ~c!}
\nonumber
\end{eqnarray}
where $m=(a+b-c)/2$, $n=(b+c-a)/2$, $p=(c+a-b)/2$. 


\vspace{0.7cm}
\noindent {(4)} The Tetrahedral net
\begin{eqnarray}
\Tet[A,B,E;C,D,F] 
&=& \CBOX[1]{\FIGThetraedron{A}{B}{E}{C}{D}{F}} \label{eq:tetraedro} \\
~~~~&=& \frac{{\cal I}}{{\cal E}} \sum_{m\leq S \leq M}
 \frac{ (-1)^{S} (S+1)!}{\prod_i  ~(S-a_i)!~\prod_j ~(b_j-S)! }
~~,
\nonumber
\end{eqnarray}
where
$$
\Ds{\begin{array}{rclcrcl} 
   a_1&=&\Ds{\frac{A+D+E}{2}},
   &\qquad&b_1&=&\Ds{\frac{B+D+E+F}{2}}, 
\\[2 mm]
   a_2&=&\Ds{\frac{B+C+E}{2}},
   &\qquad&b_2&=&\Ds{\frac{A+C+E+F}{2}}, 
\\[2 mm]
   a_3&=&\Ds{\frac{A+B+F}{2}},
   &\qquad&b_3&=&\Ds{\frac{A+B+C+D}{2}}, 
\\[2 mm]
   a_4&=&\Ds{\frac{C+D+F}{2}},&\qquad&     & &      
\end{array}}
$$
$$
\Ds{\begin{array}{rclcrcl} 
   m &=&{\rm max}\{ a_i \},   &&      
   M &=&{\rm min}\{ b_j \}, \\[2 mm]
   {\cal E} &=& A! ~B! ~C! ~D! ~E! ~F!,
 &&{\cal I} &=& \prod_{ij} (b_j-a_i)! 
~. 
\end{array}}
$$

\vspace{0.7cm}
\noindent {(5)} The Reduction Formulae

\begin{equation}
 \bra{\CBOX[3]{\Prim}} = \frac{\theta(p,p,2)}{\Delta(p)} \, 
\bra{\CBOX[2]{\Threelin}} \label{eq:upterm}
\end{equation}

\begin{equation}
 \bra{\CBOX[3]{\Terc} \:} =  
\frac{ \Tet[p,p,r;q,q,2]}{\theta(p,q,r)}\, \bra{\CBOX[2]{\Threelin}}
\label{eq:tanterm}
\end{equation}

\vspace{0.7cm}
\noindent {(6)} The recoupling theorem:
\begin{eqnarray}
\CBOX[1]{\RECOUPLINGuno{j}{a}{b}{c}{d}} 
    &=& \sum_i \SixJ[a,b,i;c,d,j] 
               \CBOX[1]{\RECOUPLINGdue{i}{a}{b}{c}{d}}
\label{eq:recTheorem}\\
\SixJ[a,b,i;c,d,j] &=& \frac{ \Ds{ \Delta_i ~\Tet[a,b,i;c,d,j]}
}{\theta(a,d,i) \theta(b,c,i) }
~. 
\label{eq:teorec}
\end{eqnarray}


\Figures


\end{document}